# Ontology-Based Recommendation of Editorial Products

Thiviyan Thanapalasingam[1], Francesco Osborne[1],
Aliaksandr Birukou[2], Enrico Motta[1]

[1] Knowledge Media Institute, The Open University, MK7 6AA, Milton Keynes, UK
`{thiviyan.thanapalasingam,francesco.osborne,enrico.motta}@open.ac.uk`
[2] Springer-Verlag GmbH, Tiergartenstrasse 17, 69121 Heidelberg, Germany
`aliaksandr.birukou@springer.com`

**Abstract.** Major academic publishers need to be able to analyse their vast catalogue of products and select the best items to be marketed in scientific venues. This is a complex exercise that requires characterising with a high precision the topics of thousands of books and matching them with the interests of the relevant communities. In Springer Nature, this task has been traditionally handled manually by publishing editors. However, the rapid growth in the number of scientific publications and the dynamic nature of the Computer Science landscape has made this solution increasingly inefficient. We have addressed this issue by creating Smart Book Recommender (SBR), an ontology-based recommender system developed by The Open University (OU) in collaboration with Springer Nature, which supports their Computer Science editorial team in selecting the products to market at specific venues. SBR recommends books, journals, and conference proceedings relevant to a conference by taking advantage of a semantically enhanced representation of about 27K editorial products. This is based on the Computer Science Ontology, a very large-scale, automatically generated taxonomy of research areas. SBR also allows users to investigate why a certain publication was suggested by the system. It does so by means of an interactive graph view that displays the topic taxonomy of the recommended editorial product and compares it with the topic-centric characterization of the input conference. An evaluation carried out with seven Springer Nature editors and seven OU researchers has confirmed the effectiveness of the solution.

## 1 Introduction

Major academic publishers need to be able to analyse their vast catalogue of editorial products and make data-driven decisions to ensure they are showcasing the right products to the right target market. This is a complex exercise that requires characterising with a high precision the topics of thousands of books and matching them with the interests of the relevant scientific communities.

In Springer Nature, this task has traditionally been handled manually by publishing editors, who tend to rely on their domain knowledge and their personal experience for selecting the books to be marketed at scientific venues. In addition to this, they typically use Springer.com[1] for searching publications associated with keywords relevant to the conferences in question and find additional information by querying their internal database of editorial products. This approach lacks a user-friendly interface and can be very time-consuming, since it requires editors to manually browse a large and fast-growing catalogue of publications. For example, in order to select books for the International Semantic Web Conference one might want to search for all the publications produced in the last three years that have been authored by well-known researchers who are likely to attend the event. While the editorial products are tagged with product market codes characterizing their topics, these are only limited to high-level research fields, such as "Artificial Intelligence" and "Database Systems". The results of the editor queries may thus include hundreds of items. Another issue is that keyword-based queries do not take in consideration the relationships between topics and may miss pertinent publications that do not contain specific strings. For instance, searching all books about "ontology matching" may miss publications about "ontology alignment".

---
[1] http://www.springer.com/

In this paper, we present Smart Book Recommender (SBR)[2], an ontology-based recommender system developed by The Open University (OU) in collaboration with Springer Nature (SN) for supporting their Computer Science editorial team in selecting products to market at specific venues. SBR recommends books, journals, and proceedings by taking advantage of a semantically enhanced representation of about 27K editorial products. In order to do so, we characterized all SN publications according to their associated research topics by exploiting the Computer Science Ontology (CSO), a large-scale automatically generated taxonomy of research areas [1]. Furthermore, SBR allows users to investigate why a certain publication was suggested by means of an interactive graph view that compares the topics of the suggested publication with those characterizing the input conference.

The rest of the paper is organized as follows. In Section 2, we discuss Smart Book Recommender in terms of its knowledge base, its architecture, and its user interface. In Section 3, we present the results of the user study. In Section 4, we discuss the steps required for large-scale deployment of the technology within the company. In Section 5, we review the state of the art and in Section 6 we conclude outlining future directions of research and development.

## 2   Smart Book Recommender

Smart Book Recommender takes as input a conference series and returns a list of editorial products that may be of interest for the attendees of the conference. This is achieved by representing SN books as a set of research topics drawn from a large-scale Computer Science ontology, and ranking them according to their similarity with a topic-centric characterization of the conference. For instance, given the conference series "International Semantic Web Conference" (ISWC), SBR will return the books, journals, and conference proceedings that are characterized by a set of research topics similar to the one of ISWC, e.g., the "Handbook of Semantic Web Technologies" and "Proceedings of the European Semantic Web Conference". The primary purpose of SBR is to provide a concise and relevant list of publications that editors can quickly review to decide which books to market during a conference. However, it can also be used by researchers for finding publications relevant to a certain venue of interest.

---

[2] A demo of SBR is available at http://rexplore.kmi.open.ac.uk/SBR-demo.

SBR provides the web interface shown in Figure 1. It works according to three main steps:

1) It represents journals, books, and conferences according to the metadata of their chapters/articles and uses the Smart Topic API [2] to characterize each of them with a semantically enhanced topic vector.
2) It computes the similarity between conferences and other editorial products and saves the results in a database.
3) For a given input conference, it returns a list of relevant editorial products, ranked by their topic-centric similarity with the conference in question and filtered in accordance with a number of user preferences.

In order to make it easier for users to understand why a certain item was suggested, SBR offers also an interactive graph view that displays the topic taxonomy of the suggested editorial product and compares it with the input conference.

In the next sections, we will discuss the system in detail. In Section 3.1, we describe the knowledge bases used by SBR. In Section 3.2, we discuss the Smart Topic API, a service for tagging books with a set of relevant topics. In Section 3.3, we describe how we compute the similarity scores. Finally, in Section 3.4, we present the user interface.

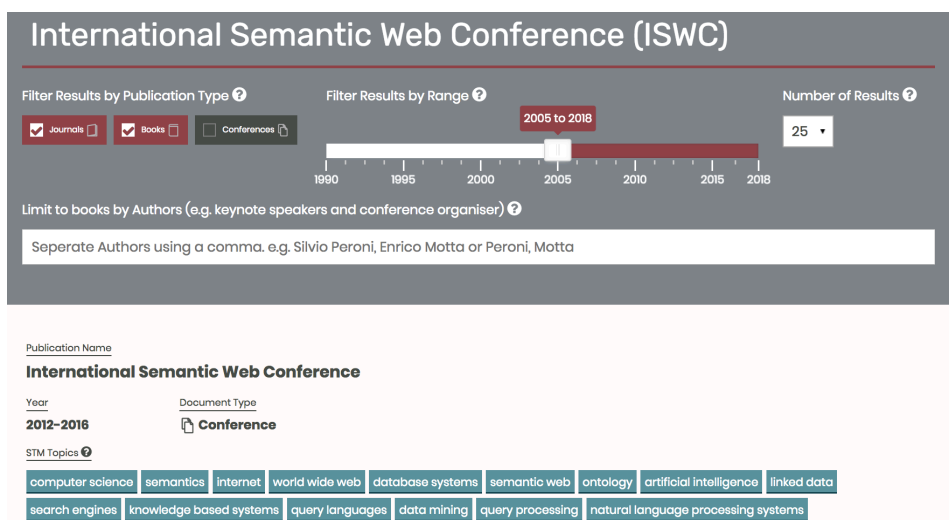

**Figure 1**. The main interface of SBR.

## 2.1 Background data

SBR relies on two background knowledge bases: a large database of metadata describing publications and the Computer Science Ontology [3].

The database of metadata contains titles, abstracts, keywords and other information describing the chapters of about **27K books and 320 journals** published by SN in the field of Computer Science. In the case of conference proceedings, journals, and edited books, each chapter is usually a research paper. Since we represent conferences according to their proceedings, SBR can only take as input conferences published by Springer Nature.

---
[3] http://skm.kmi.open.ac.uk/cso

The Computer Science Ontology (CSO) [3] is a large-scale and granular ontology of research topics that was created automatically by running the Klink-2 algorithm [1] on the Rexplore dataset [4]. This consists of about 16 million publications, primarily in the field of Computer Science. The Klink-2 algorithm combines semantic technologies, machine learning, and background knowledge from a number of web sources, including DBpedia, calls for papers, and web pages, to identify research topics and their relationships from a given corpus of publications. CSO uses the Klink data model[4], which is an extension of the BIBO ontology [5], which in turn builds on SKOS[6]. This model includes three classes of semantic relations: *relatedEquivalent*, which indicates that two topics can be treated as equivalent for the purpose of exploring research data; *skos:broaderGeneric/skos:narrowerGeneric*, which indicate that a topic is a super-area/sub-area of another one; and *contributesTo,* which indicates that the research outputs of one topic significantly contribute to the research work within another. The version of CSO used in the current prototype consists of approximately 15K semantic topics linked by 70K relationships.

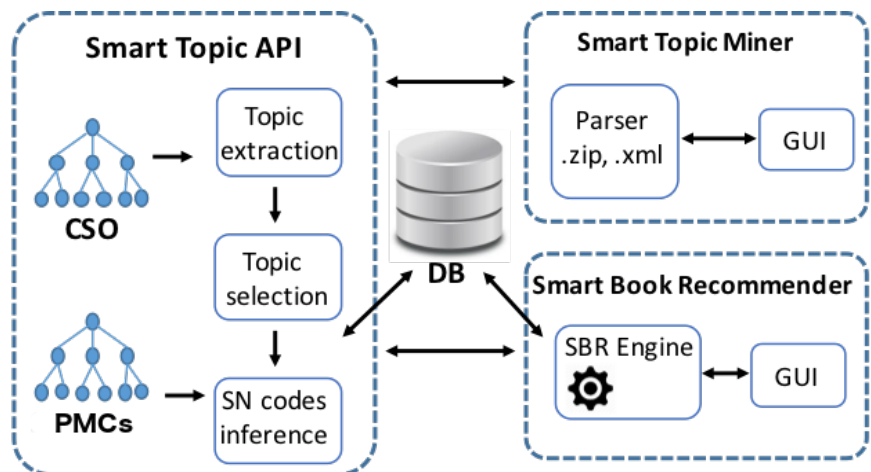

**Figure 2**. The Smart Topic API architecture.

## 2.2 The Smart Topic API

The ongoing collaboration between The Open University and Springer Nature has produced several semantic solutions for supporting the SN editorial team. These include the Smart Topic API [2, 5], an online service for automatically tagging publications with a set of relevant topics from CSO. This API supports a number of applications, including Smart Book Recommender, Smart Topic Miner [5], the Technology-Topic Framework [6], a system that forecasts the propagation of technologies across research communities, and the Pragmatic Ontology Evolution Framework [7], an approach to ontology evolution that is able to select new concepts on the basis of their contribution to specific computational tasks.

---

[4] http://technologies.kmi.open.ac.uk/rexplore/ontologies/BiboExtension.owl
[5] http://purl.org/ontology/bibo/
[6] http://www.w3.org/2004/02/skos/

Figure 2 shows the architecture of the system. The Smart Topic API takes as input a JSON containing the metadata of a book and returns its description in terms of a taxonomy (or optionally a list) of topics, in which each topic is associated with the number of chapters in which it appeared. It works as following:

1) For each topic in CSO (e.g., Semantic Web), it associates all the chapters that contain the label of the topic or the label of any *relatedEquivalent* or *skos:narrowerGeneric* (e.g., Linked Data) in the title, the abstract, or the keyword field.
2) It reduces the list of topics associated with a book to a user-friendly number by means of set covering algorithms [5].
3) It infers from the topics the product market codes (PMCs) used by SN as internal classification. It then returns a taxonomy of research topics and PMCs associated with the (number of) chapters in which they were detected.

The Smart Topic API powers Smart Topic Miner (STM) [5], a web interface that supports SN editors in classifying proceedings. STM allows editors to submit one or more proceedings, uses the API to annotate them, and then displays them as a taxonomy of research topics. It also offers a number of other options, such as the ability of explaining why a certain topic is relevant by showing the full set of sub-topics that were used to infer it. STM halves the time needed for classifying proceedings from 20-30 to 10-15 minutes and allows this task to be performed also by assistant editors, thus distributing the load and reducing costs [5] .

```
Data: List of publications SN_dataset
Result: Pairwise Similarity Values pairwiseSimilarityScores
1 foreach conference in SN_dataset.getConferences() do
      /* Extract topic distributions the proceedings of the last 5 years with the Smart Topic API */
2     topicsInConference = GetTopics(conference);
3     foreach book in SN_dataset do
4         pairwiseSimilarityScores=[];
          /* Extract topic distributions of the publication with the Smart Topic API */
5         topicsInBook = GetTopics(book);
          /* Estimate the similarity using the Jaccard index */
6         jaccard_heuristic = computeJaccardIndex(topicsInConference, topicsInBook);
7         if jaccard_heuristic >= JaccardIndexThreshold then
8             cosine_similarity = computeCosineSimilarity(topicsInConference, topicsInBook);
9             if cosine_similarity >= SimilarityThreshold then
10                pairwiseSimilarityScores[] = {conference, book, cosine_similarity};
11            end if
12        end if
13    end foreach
      // Save set of similarity scores to SBR database
14    storeInDatabase( pairwiseSimilarityScores );
15 end foreach
```

**Algorithm 1**. The SBR algorithm

### 2.3 Similarity Computation

In order to characterize specific journals, books, and conferences we group the publications as following: 1) for books, chapters are grouped by the book DOI; 2) for journals, the articles are grouped using the journal DOI and their publication year (e.g.,

Journal of Intelligent Information Systems in 2016), and 3) for conferences, papers are grouped using unique conference identifiers and considering only articles from the last five years. We use the persistent identifiers for conferences and conference series introduced in the Linked Open Data Conference Portal [8] and recently migrated to SciGraph [9]. Such identifiers make sure that the conference series links all relevant conferences, regardless of name changes (e.g., after a few years the "European Semantic Web Conference" became the "Extended Semantic Web Conference") and acronyms.

In an earlier version of SBR, we considered specific editions of conferences –e.g., ISWC 2013. However, on the basis of feedback from the editors, it was decided to consider full conferences series rather than individual editions. This solution simplifies the interface and allows us to reduce possible bias from specific conference editions, which may be affected by trendy topics exhibiting a transient burst of popularity.

We employ the Smart Topic API to associate each item with a vector in which the elements represent research topics and their value is the number of chapters/papers in which the topic was detected. Henceforth, this value will be referred to as *topic weight*.

We exploit this vector representation for computing the similarity between the conference series and the editorial products, as described in Algorithm 1. Since, the Smart Topic API associates publications containing topic T also with the *relatedEquivalent* and *skos:broaderGeneric* of T (see S2.2), the resulting vectors allow us to match publications that refer to the same concepts (e.g., "Deep Learning") with different key phrases ("Deep Neural Networks"), at different granularity levels ("Machine Learning", "AI").

We assess the similarity of two semantic vectors using the cosine similarity [10], since this measure relies on the orientation but not the magnitude of the topic weights in the vector space, allowing us to compare editorial products associated with a different number of chapters. The similarity computation is carried out offline.

Since it is computationally infeasible to calculate the cosine similarity between each book in the SN dataset, we first prune the number of candidate pairs by calculating their Jaccard index, which is a more lightweight similarity metric, and selecting only the ones that yield a value higher than a threshold. A data analysis revealed that by applying a threshold of 0.125 we halve the number of candidate pairs while still producing very good results. Finally, we save the cosine similarity of a pair in the database if it is greater than 0.5, since according to the editors, recommendations with similarity < 0.5 are unhelpful.

### 2.4 The web interface

Figure 1 shows the user interface of SBR. The user can select a conference by typing either the conference name (e.g., "International Semantic Web Conference") or its abbreviated form (e.g., "ISWC"). In Figure 1 the user has selected ISWC and SBR is showing the top fifteen topics that characterize this venue.

When the user selects a conference, the corresponding conference ID along with the other user preferences (e.g., publication type, year, maximum number of results) are sent as JSON file to the backend via a GET request. The backend is a REST API, which retrieves all relevant publications that meet the criteria and returns the results as a JSON file, which is then visualized by the web interface. The API was developed in Python and the data are pulled from a MariaDB database, while the frontend uses HTML5 and Javascript.

Here, we briefly describe the settings available to the users, to allow them to customise the behaviour of the system.

- *Types of publication* – Users can specify which types of editorial products should be included in the results. Currently, these include books, journals, and (other) conference proceedings.
- *Publication year* –Users can filter results to include only the ones published in a specified time interval. By default, this interval is set to the last three years.
- *Maximum number of results* – Users can set the maximum amount of results according to their needs. This functionality is provided as normally editors can only select a limited number of books to market during a conference.
- *Filter publications by authors and editors* – Users can narrow down the recommendations to books authored or edited by an individual or a group of academics using this free text field. This functionality is provided as editors often focus on marketing editorial products produced by key researchers with high visibility in the research fields relevant to a conference.
- *Exporting data* – Once a list of recommendations has been generated, it is possible for the user to export the results as a CSV or JSON file. These files are typically used by publishing editors to submit an order to the Exhibit Department, which takes care of dispatching the selected products to the conference.

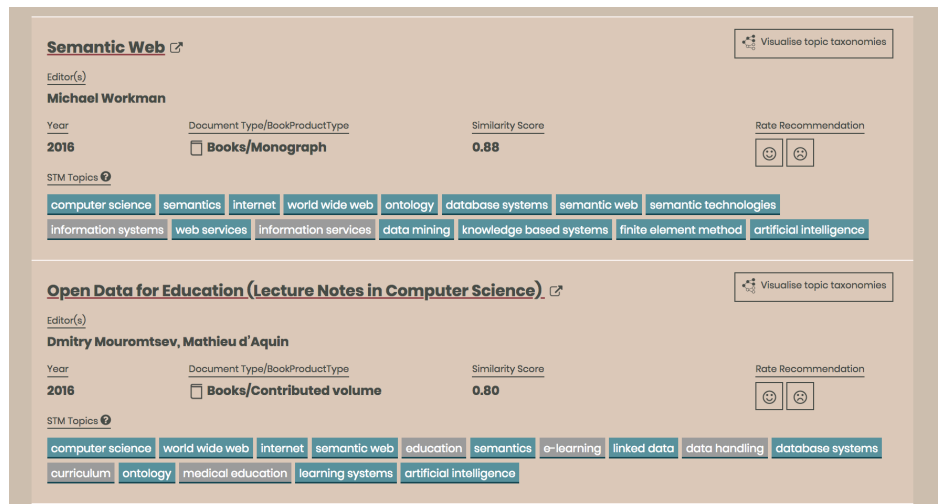

**Figure 3**. Recommended SN books for ISWC.

Figure 3 shows the recommendation list that is loaded via an AJAX request after the user has selected a conference. The results are shown as *cards* and sorted in descending order of similarity. Each publication is summarized with respect to its key elements. These include title, publication year, the fifteen most significant topics, and the overall similarity score with the input conference. We display the authors of a book wherever there are less than five authors, otherwise we display editors.

The users can interact with each card by:

- *Examining the publication on SpringerLink* [7] – A hyperlink on the publication title redirects users to the relevant SpringerLink page. This enables editors to collect additional information regarding the publication, such as the authors of individual papers and the abstracts.
- *Providing feedback for a specific card* – We provide a binary feedback system that uses emoticon radio buttons to allow users to express their view on a recommendation. The feedback is used to improve the recommender engine.
- *Opening the graph view interface* – By clicking on the "visualize topic taxonomy" button, users can access a graph view, shown in Figure 4, which makes it easier for them to make sense of the relationship between the selected output and the input conference.

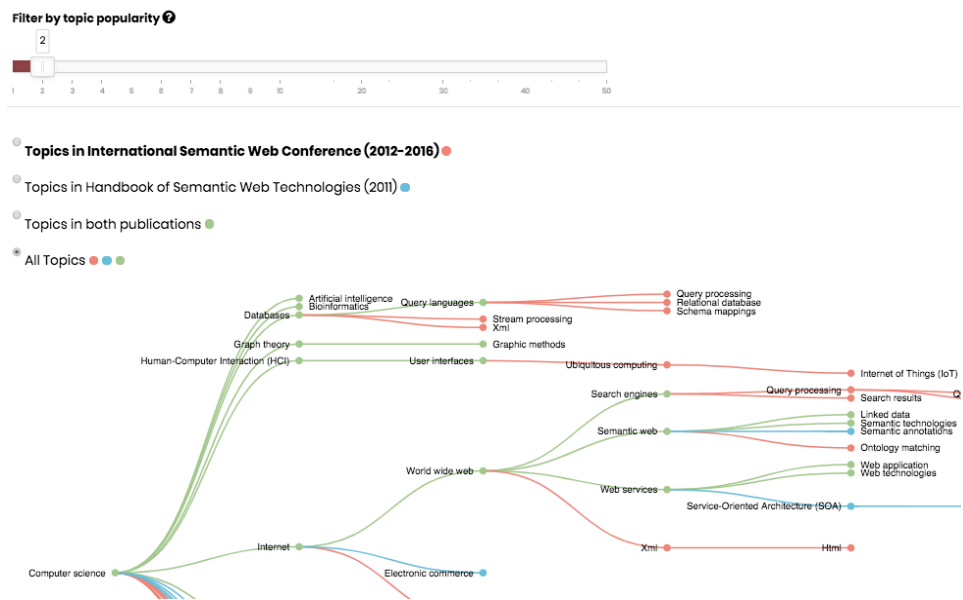

**Figure 4**. Portion of the graph view showing the taxonomies of the topics associated with the input conference and one of the recommended editorial products.

The graph view [8] visualizes editorial products according to their taxonomy of research topics derived from the Computer Science Ontology. The purpose is allowing users to understand why a certain product was recommended and how its associated topics intersects with the ones characterizing the input conference. As an example, in Figure 4 we show the comparison between the topic-centric characterization of the "*Handbook of Semantic Web Technologies*"[9] and the one of the International Semantic Web Conference. The user can choose whether to visualise only the topics of the conference, those of the recommended publication, their intersection, or all topics of the two items. Hovering over a topic shows the number of chapters/papers within the publication that are associated with the topic. A slider above the interface allows users to filter topics according to their weight.

---

[7] https://link.springer.com/
[8] The graph view was realised in JavaScript, using the D3.js library (https://d3js.org/).
[9] https://link.springer.com/referencework/10.1007%2F978-3-540-92913-0

## 3 Evaluation

We evaluated Smart Book Recommender by means of a user study involving seven SN editors and seven OU researchers. The goal of the study was to assess both the usability of SBR and also the quality of its recommendations. We structured the user study in three phases. First, we provided each subject with a 10 minute introduction to SBR. Then we asked them to try the system for approximately 45 minutes and rate its recommendations. Finally, each subject filled a questionnaire about their experience with SBR.

| *Option* | *Applies to* | *Definition* |
| --- | --- | --- |
| Bring it | Editor only | The item is *relevant* to the conference and the editor would *bring* this item to the conference and market it. |
| Read it | Researcher only | The item is *relevant* to the conference and the researcher would want to *read it*. |
| Relevant | Both | The item is *relevant* to the conference, but the editor does not consider it suitable to be marketed or the researcher does not desire to read it. This could be for a variety of marketing or personal reasons. |
| Debatable | Both | Whether the recommended item is relevant to the conference is open to discussion and different people may have different opinions. |
| Irrelevant | Both | The recommended item is not relevant to the conference and should not be recommended. |

**Table 1**. Options available to editors and researchers for rating recommendations.

While editors are the main users of the system, we also evaluated SBR with a number of researchers, given that the whole point of the application is to assist editors in selecting editorial products that researchers are likely to be interested in. The expertise of the evaluators covered a variety of Computer Science topics, including but not limited to Robotics, Semantic Web, Software Engineering, HCI, AI, Computational Biology, and Wireless Networks.

We assessed the quality of the results by considering the "bring it", "read it", and "relevant" books as relevant instances and computing the Precision @10, a standard metric for evaluating ranked lists of items.

The material produced for this evaluation is publically available at http://rexplore.kmi.open.ac.uk/SBR_eval_data and on FigShare [10].

### 3.1 Quantitative Analysis

We assessed the performance of SBR in suggesting relevant publications, by asking users to choose two conferences in their fields of expertise and then rate SBR recommendations. For each conference, SBR suggested 20 books and 10 conference proceedings. To keep recommendations consistent, we considered all books and proceedings published between 2005 and 2018, regardless of the authors and editors.

---

[10] https://doi.org/10.6084/m9.figshare.6087032.v2

We asked the users to rate each item by selecting one of the four options presented in Table 1. The sessions were video-recorded to allow further analysis.

The Precision @10 is 76.8% for the books and 75.4% for the proceedings. It thus seems that there is not much difference in the quality of the recommendations regarding these two editorial products.

Figure 5 shows the percentage of recommendations that were tagged as "bring it/read it", "relevant", "debatable", or "irrelevant" by the users. Editors rated "bring it" or "relevant" 72.9% of the recommendations while researchers rated "read it" or "relevant" the 66.8% of them. In total, only 10.5% of the recommendations were rated as irrelevant by the editors and 7.2% by the researchers.

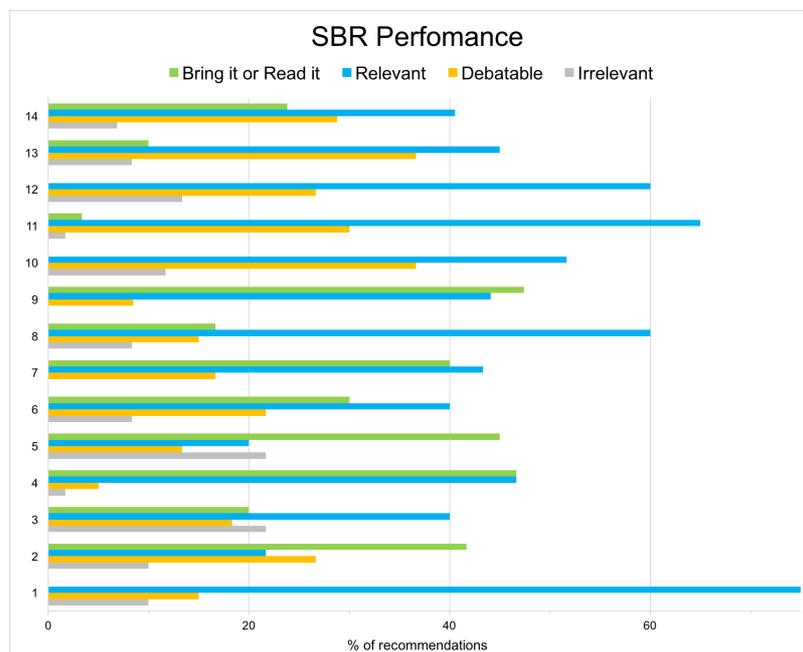

**Figure 5.** SBR performance as rated by the evaluators (SN editors labelled 1-7 and OU researchers 8-14). The results of the 28 test cases were aggregated by user.

Editors would bring to the conference 31.9% of the recommended publications, considering the others not marketable for a variety of reasons, even when they were relevant. On the other hand, researchers would read 14.5% of it. This discrepancy may be explained by the fact that editors and researchers apply different decision-making strategies when choosing whether to "bring" or "read" a publication. Researchers are mainly interested in publications that address their specific needs and they consider also the time invested in reading it and the price. Conversely, editors take into account the preferences of a large group of people and consider a variety of other dimensions, such as how much the book sold in previous years, the popularity of the authors within the community, the potential audience size, and so on.

## 3.2 Qualitative Analysis

The questionnaire consisted of three sections: i) an assessment of the evaluators' background and expertise, ii) five open questions, and iii) a standard System Usability Scale (SUS) questionnaire to assess the usability of SBR. On average, the editors had 15 years of experience in their role and extensive experience in selecting books for conferences. Three of them had more than 20 years of experience in their field. The OU researchers had an average seniority of about 5 years.

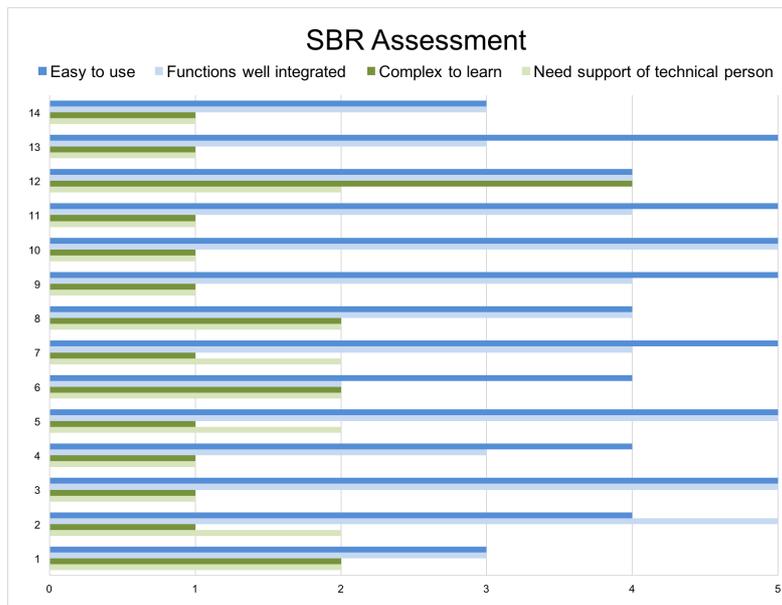

**Figure 6.** SUS questionnaire results (editors labelled 1-7 and researchers 8-14).

We will first summarize the answers to the open questions.

**Q1. How do you find the interaction with the SBR interface?** Both groups found the user interface very intuitive. Most attributed this to the "simple" and "well-organised" layout of SBR and the ability to perform queries with little user input. One researcher mentioned that there was a learning curve but it was "easy to pick up", and one editor suggested to make the text input field for searching conference series more noticeable.

**Q2. How effectively does SBR support you in selecting relevant publications?** Some editors placed the accuracy of the recommended conference proceedings higher than that of the books. One editor felt that some recommended titles were generic, possibly due to the "large margin of error associated with vast selections of conferences" and two pointed out that it would be beneficial to be able to select particular book types, such as handbooks, textbooks and monographs.

**Q3. What are the most useful features of SBR?** Five researchers found the visual analytics of taxonomies useful for understanding similarities. Three editors appreciated the hyperlinks to the Springer product page. Two researchers and one editor found particularly helpful the option of viewing books and conferences independently.

**Q4. What are the main weaknesses of SBR?** There was general agreement between editors and researchers that supporting only Springer published conferences is a significant drawback. Three editors indicated that some of the book titles relevant to their conference were not recommended. Another two mentioned that when searching for books, the system returned also some proceedings (i.e., books from the LNCS series).

**Q5. Can you think of any additional features to be included in SBR?** Two researchers and two editors would like to have the ability of modifying the automatic representation of the input conference by adding or removing some topics. Some editors would like to have direct links to conference pages and additional information about publications, e.g., the main subject discipline and whether they are open access or not.

The last part of the user survey consisted of a SUS questionnaire, a standard tool for assessing the usability of an application. The SUS questionnaire includes 10 questions on a 1 to 5 scale, where 1 is the most negative assessment and 5 the most positive. The average system is expected to score 68 out of 100. The editors and the researchers yielded respectively an average SUS score of 77.1±15.2 and 80.3±11.3, which converts in a percentile rank of about 75%.

Figure 6 shows the answers of the users to four SUS questions. The users believed that SBR was easy to use (with an average score of 4.4±0.7) and its functions where well integrated (3.9±0.6). They did not think that it was complex to use (1.4±0.8) or that they would need the help of a technical person to use it in the future (1.5±0.5).

### 3.3 Informal feedback beyond Computer Science editorial team

In addition to the formal evaluation reported in this section, we have also presented the SBR tool to a wider group of publishing editors and editorial assistants at Springer Nature. The fifteen participants (3 sessions with 5 participants each) first saw a short 3-minute demo of the tool and then took part in a 10-12 minute session where they were encouraged to ask questions and suggest improvements.

The participants saw strong potential for the SBR tool over current practices, which include "ask colleagues for relevant books and journals via e-mail, hoping they have time to reply and are in the office" and "compile a list of relevant titles using various systems, actually developed for other purposes". In particular, they appreciated the time range and type of product filters and the support for searching for books by keynote speakers. They also suggested areas for further improvements, such as the ability of i) directly querying the system with a list of research topics; ii) looking up people on the editorial board of a journal; iii) expand the scope of the system to other disciplines (e.g., Mathematics).

### 3.4 Discussion and Limitations

SBR obtained a more than satisfactory performance in recommending relevant editorial products and received a high score in term of usability. Nonetheless, the evaluation highlighted some issues that we intend to address in future versions.

A first concern that was mentioned by a number of users is that SBR currently provides recommendations only for conferences which proceedings are published by

Springer Nature[11], thus not providing support for marketing activities outside these conferences. In order to include more conferences, we need to also access to the conference proceedings published by other editors. We are thus exploring the option of using datasets such as CrossRef [12], Dimensions [13], OpenCitations [11], and Core [12].

Another issue arising from the evaluation is that sometimes the topic characterization of books with few chapters is quite sparse. In these cases, considering only title, abstract and keywords may not allow to identify enough topics to allow a fair comparison with the other editorial products. We are thus considering using also the full text.

A third issue that emerged during the evaluation is the coverage of multidisciplinary publications. SBR represent topics by means of the Computer Science Ontology, and therefore scarcely covers other fields such as Biology, Engineering, Mathematics, or Economics. Therefore, publications which include other fields in addition to Computer Science are sometimes misrepresented, lowering the overall quality of the recommendations. We plan to address this issue by applying the ontology learning techniques utilized to produce the Computer Science Ontology also on other domains of science.

Finally, some users mentioned that they would like the option of modifying the set of topics that get extracted from the conference proceedings and is used to produce the recommendations. A further step in this direction would be to allow users to input directly a set of topics as a query. This would naturally require some significant changes to the backend, since currently all the similarity values are precomputed, but it would also allow for more flexibility. Indeed, this solution may also enable us to associate users with a representation of their research interests and automatically produce tailored recommendations.

## 4  Next steps for large scale deployment

SBR was well received by Springer Nature editors, but we must take some additional steps to fully integrate it into their workflow.

In the first instance, we intend to automatize the process for importing and processing the most recent editorial items. Currently, we renew our database every four months by importing a new dump of metadata and recalculating the similarity values. This solution suffers from two limitations: it requires human intervention and the system is updated only every four-months. We plan to fully automatize this process by developing a system for importing new metadata on a daily basis and recomputing seamlessly the relevant similarity values.

In the second instance, we plan to develop a new version of SBR that will address the most important requests that came up during the user study, as discussed in previous section.

Finally, we are exploring the ability of SBR to produce collections of documents relevant to certain topics, e.g., all recent publications in the field of Ontology Engineering. This has broader implications beyond selecting books for conferences, and can help compiling ad-hoc packages for industry or academic institutions in the developing countries. Some initial experiments in this direction have already yielded promising results.

---

[11] Actually, since Springer Nature is one of the largest publishers of Computer Science conferences, its coverage of the conferences in this field is very extensive.
[12] http://crossref.org
[13] https://www.dimensions.ai

## 5  Related Work

Recommender systems are software tools and methods which provide suggestions for items to users, according to their preferences and needs [13]. They are typically classified as collaborative filtering approaches, content-based filtering approaches and hybrid approaches [14].

Content-based recommender systems [15] rely on a pre-existing domain knowledge to suggest items more similar to the ones that the user seems to like. They usually generate user models that describe user interests according to a set of features [16]. With the advent of the Semantic Web, several recommender systems started to adopt ontologies for representing both user interests and items [17]. Often these systems use an ontology so that, given user interest in an item represented in the ontology, they can then propagate such interest to relevant items and concepts. For example, given a positive feedback on "beagles", a system may infer (correctly or not) that a user is interested in "dogs", and more generally in "pets". SBR exploits a similar mechanism when it infers that a publication explicitly linked to a topic (e.g., Linked Data) is also about its *skos:broaderGeneric* concepts in CSO (e.g., Semantic Web). The main advantages of these solutions are i) the ability to exploit the domain knowledge for improving the user modelling process, ii) the ability to share and reuse system knowledge, and iii) the alleviation of the cold-start and data sparsity problems [16, 18].

We will now discuss some of these ontology-based approaches. Sieg et al. [16] present an ontology-based recommender to improve personalised Web searching in which the user profiles are instances of a reference domain ontology and are incrementally updated based on the user interaction with the system. Middleton et al. [18] describe a hybrid recommender system that exploit ontologies for increasing the accuracy of the profiling process and hence the usefulness of the recommendations. Thiagarajan et al. [19] use a different strategy by representing user profiles as bags-of-words and weighing each term according to the user interests derived from a domain ontology. Razmerita et al. [20] describe OntobUM, an ontology-based recommender that integrates three ontologies: i) the user ontology, which structures the characteristics of users and their relationships, ii) the domain ontology, which defines the domain concepts and their relationships, and iii) the log ontology, which defines the semantics of the user interactions with the system. Birukou et al [21] present an agent-based system that learns the preferences of experienced researchers and provides specific suggestions to support search for scientific publications. Colombo-Mendoza et al [22] propose RecomMetz, a context-aware mobile recommender system based on Semantic Web technologies. This system introduced some unique features, such as the composite structure of the items and the integration of temporal and crowd factors into a context-aware model. Finally, Cantador et al. [23] propose a hybrid recommendation model in which user preferences are described in terms of semantic concepts defined in domain ontologies. Similar to all these systems SBR builds a semantic representation of the items and exploits the ontology for inferring additional concepts. However, rather than creating a representation of a single user, it characterizes the overall interests of the research community associated with the proceedings of a conference series.

SBR builds on the Smart Topic API to represent publications as vectors of research topics. This is a useful representation that is used in a variety of systems for exploring the research landscape [4, 6]. In recent years, we have seen the emergence of several approaches to annotating research articles. For instance, DBpedia Spotlight [24] is often used for automatically annotating papers with DBpedia concepts. Gabor et al. [25] introduce an approach for annotating scientific corpora with domain-relevant concepts and semantic relations. The Dr. Inventor Framework [26] focuses instead on extracting structured textual contents, discursive characterization of sentences, and graph based representations of text excerpts.

## 6  Conclusions

In this paper, we presented Smart Book Recommender, a semantic recommender system developed in collaboration with Springer Nature which suggests editorial products to market at academic venues.

A user study involving seven SN editors and seven OU researchers showed that SBR was able to suggest relevant materials and scored high in usability. In particular, Springer Nature editors considered as relevant 72.9% of the SBR recommendations and assessed the system as very user friendly, yielding an average SUS score of 77.1.

We are now planning to further integrate the SBR tool into the process workflows at Springer Nature. To this purpose, we are going to develop a new version of the system, which will take into account a variety of suggestions which arose from the user study.

## Acknowledgements

We would like to thank publishing editors at Springer Nature for assisting us in the evaluation of SBR and allowing us to access their large repositories of scholarly data.

## References

1. Osborne, F., Motta, E.: Klink-2: Integrating Multiple Web Sources to Generate Semantic Topic Networks. In: International Semantic Web Conference. pp. 408–424. Springer (2015).
2. Osborne, F., Salatino, A., Birukou, A., Thanapalasingam, T., Motta, E.: Supporting Springer Nature Editors by means of Semantic Technologies. In: International Semantic Web Conference. Springer (2017).
3. Salatino, A.A., Thanapalasingam, T., Mannocci, A., Osborne, F., Motta, E.: The Computer Science Ontology : A Large-Scale Taxonomy of Research Areas. In: International Semantic Web Conference 2018 , Monterey, CA (USA) (2018).
4. Osborne, F., Motta, E., Mulholland, P.: Exploring Scholarly Data with Rexplore. In: International Semantic Web Conference pp. 460–477. Springer, Berlin, Heidelberg (2013).
5. Osborne, F., Salatino, A., Birukou, A., Motta, E.: Automatic Classification of Springer Nature Proceedings with Smart Topic Miner. In: International Semantic Web Conference. pp. 383–399. Springer (2016).
6. Osborne, F., Mannocci, A., Motta, E.: Forecasting the Spreading of Technologies in Research Communities. In: Proceedings of the Knowledge Capture Conference (2017).

7. Osborne, F., Motta, E.: Pragmatic Ontology Evolution: Reconciling User Requirements and Application Performance. In: International Semantic Web Conference 2018 , Monterey, CA (USA). (2018).
8. Birukou, A., Bryl, V., Eckert, K., Gromyko, A., Kaindl, M.: Springer LOD Conference Portal. Demo paper. In: International Semantic Web Conference. Springer, Vienna Austria (2017).
9. Hammond, T., Pasin, M., Theodoridis, E.: Data integration and disintegration: Managing Springer Nature SciGraph with SHACL and OWL. In: Proceedings of the ISWC 2017 Posters & Demonstrations and Industry Tracks. Springer, Vienna, Austria (2017).
10. Salton, G., Buckley, C.: Term-weighting approaches in automatic text retrieval. Inf. Process. Manag. 24, 513–523 (1988).
11. Peroni, S., Dutton, A., Gray, T., Shotton, D.: Setting our bibliographic references free: towards open citation data. J. Doc. 71, 253–277 (2015).
12. Knoth, P., Zdrahal, Z.: CORE: Three Access Levels to Underpin Open Access. D-Lib Mag. 18, (2012).
13. Ricci, F., Rokach, L., Shapira, B. eds: Recommender Systems Handbook. Springer US, Boston, MA (2015).
14. Burke, R.: Hybrid Web Recommender Systems. In: The Adaptive Web. pp. 377–408. Springer, Berlin, Heidelberg (2007).
15. Lops, P., de Gemmis, M., Semeraro, G.: Content-based Recommender Systems: State of the Art and Trends. In: Recommender Systems Handbook. pp. 73–105. Springer US, Boston, MA (2011).
16. Sieg, A., Mobasher, B., Burke, R.: Web search personalization with ontological user profiles. In: Conference on Information and Knowledge Management. p. 525. ACM Press, New York, New York, USA (2007).
17. de Gemmis, M., Lops, P., Musto, C., Narducci, F., Semeraro, G.: Semantics-Aware Content-Based Recommender Systems. In: Recommender Systems Handbook. pp. 119–159. Springer US, Boston, MA (2015).
18. Middleton, S.E., Shadbolt, N.R., De Roure, D.C.: Ontological user profiling in recommender systems. ACM Trans. Inf. Syst. 22, 54–88 (2004).
19. Thiagarajan, R., Manjunath, G., Stumptner, M.: Finding Experts By Semantic Matching of User Profiles. In: International Semantic Web Conference Personal Identification and Collaborations: Knowledge Mediation and Extraction (PICKME 2008). pp. 7–18 (2008).
20. Razmerita, L., Angehrn, A., Maedche, A.: Ontology-Based User Modeling for Knowledge Management Systems. In: International Conference on User Modeling. pp. 213–217. Springer, Berlin, Heidelberg (2003).
21. Birukou, A., Blanzieri, E., Giorgini, P.: A Multi-Agent System that Facilitates Scientific Publications Search. In: International Joint Conference on Autonomous Agents and Multiagent Systems. pp. 265–272. ACM Press (2006).
22. Colombo-Mendoza, L.O., Valencia-García, R., Rodríguez-González, A., Alor-Hernández, G., Samper-Zapater, J.J.: RecomMetz: A context-aware knowledge-based mobile recommender system for movie showtimes. Expert Syst. Appl. 42, 1202–1222 (2015).
23. Cantador, I., Bellogín, A., Castells, P.: A multilayer ontology-based hybrid recommendation model. AI Commun. 21, 203–210 (2008).
24. Mendes, P.N., Jakob, M., García-Silva, A., Bizer, C.: DBpedia Spotlight: Shedding Light on the Web of Documents.
25. Gábor, K., Zargayouna, H., Buscaldi, D., Tellier, I., Charnois, T.: Semantic Annotation of the ACL Anthology Corpus for the Automatic Analysis of Scientific Literature.
26. Ronzano, F., Saggion, H.: Dr. Inventor Framework: Extracting Structured Information from Scientific Publications. Presented at the (2015).